\documentclass[preprint,12pt]{elsarticle}

\usepackage{lineno,hyperref}
\usepackage{graphicx}
\usepackage{multicol}
\usepackage{amsmath,amssymb}
\usepackage{xcolor}
\usepackage{float}
\begin{document}

\begin{frontmatter}

%\title{Elsevier \LaTeX\ template\tnoteref{mytitlenote}}
\title{Enhanced attraction between particles in a bidisperse mixture with random pair-wise interactions}
\author{Madhu Priya}
\ead{madhupriya@bitmesra.ac.in}
\address{Department of Physics, Birla Institute of Technology Mesra, Ranchi -- 835215, Jharkhand, INDIA}

\author{Prabhat K. Jaiswal\corref{mycorrespondingauthor}}
\cortext[mycorrespondingauthor]{Corresponding author}
\ead{prabhat.jaiswal@iitj.ac.in}
\address{Department of Physics, Indian Institute of Technology Jodhpur, Karwar -- 342037, Rajasthan, INDIA}

\begin{abstract}
Motivated by growing interests in multicomponent metallic alloys and complex fluids, we study a complex mixture with bidispersity in size and polydispersity in energy.  The energy polydispersity in the system is introduced by considering random pair-wise interactions between the particles. Extensive molecular dynamics simulations are performed to compute potential energy and neighborhood identity ordering (NIO) parameter as a function of temperature for a wide range of parameters including size-ratio and concentration of the two species by quenching it from a high temperature fluid state to a crystalline state. Our findings demonstrate an enhancement of the neighborhood identity ordering on addition of particles of different sizes. Moreover, a comparatively higher increase in NIO parameter is achieved by tuning the size-ratio of the particles. We also propose NIO parameter to be a good marker to differentiate systems (below the liquid-to-solid transition temperature) having different values of size-ratio and concentrations. Effect of cooling rates on NIO parameter is also discussed.
\end{abstract}


\end{frontmatter}

%\linenumbers

\section{Introduction}
\label{sec1}
Understanding the relation between microscopic structure of a material and its response to mechanical perturbations is of great technological and industrial significance. Designing materials with desired properties and strength have led scientists to consider alloys with many components which may vary in interactions, shapes, sizes, etc. \cite{murty14, ms17, cantor04, cantor14, gludovatz14, nnh12}. For developing a new material, traditional alloying strategy is to select one dominant component and add other elements in small amounts to improve specific properties. Such techniques put restriction on further improvement of mechanical properties; and while achieving high strength ($\sim$ GPa), usually failure of materials occurs. To explore even wider range of remarkable new materials, strategies like equiatomic substitution of main element with multi-element systems have been very successful and reviewed in Ref.~\cite{cantor14}. These high-entropy alloys \cite{murty14, ms17, gylz16, oybg13} form a variety of amorphous and quasi-crystalline structures. Moreover, they can also crystallize as a single phase and exhibit good ductility, high toughness, superior strength, etc. For example, one recent study on a five component CrMnFeCoNi alloy, showing a one-phase face-centered cubic (fcc) solid solution, found tensile strength above 1 GPa with large fracture toughness \cite{gludovatz14}. Another study on a different five component FeCrMnNiCo alloy also forms a single-phase fcc solid solution and crystallizes in dendritic phase \cite{cantor04}. Recent work on AlCoCrFeNi$_{2.1}$ alloy simultaneously achieves high fracture strength and high tensile ductility at room temperature \cite{ldg14}. 

These metal alloys find applications ranging from nuclear reactors to metallic biomaterials. Due to excellent corrosion resistance at high temperatures \cite{ty14}, they are well suited for environments with extreme radiation exposure and hence, offers more stability to components of future nuclear reactors \cite{granberg16}. Grandberg et al. performed experiments and molecular dynamics simulations on NiCoCr alloy and found a significant reduction in radiation damage for equiatomic alloys as compared to the corresponding pure element Ni \cite{granberg16}. The concept of near or equiatomic compositional alloying has been also applied to bulk metallic glasses (BMG) \cite{it11}. One interesting result on high-entropy BMGs is that they can exhibit amorphous phase, crystalline phase, etc. under certain conditions, e.g., Al$_{0.5}$TiZrPdCuNi  alloy showed remarkable ability to be into a glassy phase as well as crystalline phase depending on the sample size and the cooling rate \cite{twc13}. Many new metallic alloys are being developed for various biomedical applications in implantation of hard tissues and bones \cite{nnh12}. Materials like Ca$_{20}$Mg$_{20}$Sr$_{20}$Yb$_{20}$Zn$_{20}$ have been reported to promote osteogenesis and improved mechanical properties; and have shown good resistance to corrosion \cite{lxz13}. 

The polydispersity in a multicomponent system is not only due to the distribution in size of the constituent atoms or molecules. A system can also be polydisperse due to the distribution of many other variables characterizing the system, e.g., charge, shape, interaction energy, etc. In addition to experimental works mentioned above, these systems have drawn attention of researchers from theoretical and computational perspectives \cite{sollich02, jf13, singh2019, shagolsem2015, shagolsem2016, dino2016, ih15, ih16, thm17, thsh18, srbr19, an19, gzmhcmgs20, nb20, pf19,rwps17}. The computer simulations of one such extreme case of mixing, where all particles are considered to be different by considering random interactions between them, has shown significant improvement in cohesive forces between the particles which in turn increases the yield strength of the system \cite{singh2019}. The motivation for choosing all pair-wise particle interactions to be different comes not only from high-entropy alloys but also from modeling complex fluids, e.g., foams, emulsions, colloidal assemblies, proteins, and granular materials, where each of the particles differs in size, shape, and interaction with its neighbors \cite{bb11, bdb17}. In this work, we test whether mixing particles of two sizes (i.e., bidisperse in size) can also enhance the neighborhood identity ordering (NIO) of an energy polydisperse system where all pair-wise particle interactions are different.

\section{Model and Simulation Details}
\label{sec2}
\noindent  To investigate the properties of systems with bidispersity in size and polydispersity in energy, particles with two different sizes are considered where all interactions between the particle pairs are different (AID)~\cite{singh2019,shagolsem2015}. The bigger particles are chosen as species A while the smaller particles are of type B. The concentration of the bigger particles is the ratio of the number of bigger particles $N_A$ and the total number of particles $N$, {\it i.e.}, $x_A = N_A/N$. The concentration of smaller particles, therefore, is $x_B=1-x_A$. The size disparity between the bigger particles and the smaller particles is defined in terms of size-ratio  $\alpha = \sigma_B/\sigma_A$, where $\sigma_A$ and $\sigma_B$ are like diameters of bigger and smaller particles, respectively. As discussed in Sec. \ref{sec1}, energy polydispersity is modeled by the random interactions between all the particle pairs $\epsilon_{ij}$. Here, values of $\epsilon_{ij}$ are withdrawn from a uniform distribution between 1 and 4~\cite{singh2019,shagolsem2015,shagolsem2016}, so that the mean value of $\epsilon_{ij}$ is 2.5.  All the fluid particles are considered to have mass as unity. The fluid particles interact with each other via the Lennard-Jones (LJ) potential,
\begin{equation}
U_{ij}^{ss'}(r)=4\epsilon_{ij}\Big[\Big(\frac{\sigma_{ss'}}{r}\Big)^{12}-\Big(\frac{\sigma_{ss'}}{r}\Big)^ { 6 } \Big ],
\end{equation}
where $i,j$ is the particle index and $s,s' \in A, B$. $\sigma_{ss'}$ is computed by considering the arithmetic mean, {\it i.e.}, $\sigma_{ss'} = (\sigma_s + \sigma_{s'})/2$ The potential is truncated and shifted at $r=r_{\rm cut}=2.5\sigma$, so that the 
truncated potential $\tilde{U}_{ij}^{ss'}(r)$ is defined as \cite{allen87, frenkel02},
\begin{eqnarray} \tilde{U}_{ij}^{ss'}(r) =
\begin{cases}
U_{ij}^{ss'}(r) - U_{ij}^{ss'}(r_{\rm cut}) & \quad {\text {if}}~ r < r_{\rm cut} \\
0 & \quad {\text {if}}~ r \ge r_{\rm cut} \; .\\
\end{cases}
\end{eqnarray}
The physical quantities measured here are reported in reduced LJ units \cite{allen87, frenkel02}. The temperature $T$ is expressed in units of $\epsilon/k_B$ where interaction parameter $\epsilon$ and Boltzmann constant $k_B$ are both taken as unity. The lengths are expressed in units of $\sigma_A$ which is also taken as unity. The fluid mixture is simulated in $NVT$ 
ensemble. The dynamics is solved by using a velocity-Verlet integrator \cite{verlet82} with a 
time step of 
$\delta t = 0.005 \tau_{LJ}$, where $\tau_{LJ} = \sigma_A(m/\epsilon)^{1/2} = 1$ 
is the LJ time unit. The multicomponent fluid mixture is simulated at three densities $\rho =0.7$, $0.8$, and $0.9$. We consider $N=1024$ and vary $N_A$ to investigate the system for many different concentrations of bigger particles ranging from $0$ to $1$. The particles are simulated in a two dimensional box of length $L = \sqrt{N/\rho}$. Periodic boundary conditions with period $L$ are imposed along the $x$ and $y$ directions. At time $t=0$, the fluid particles are placed on a square lattice and their velocity is chosen from a Maxwell-Boltzmann distribution at a high temperature set at $T=5$ for the simulations done here. The system is then equilibrated for $t=100 \tau_{LJ}$. A constant temperature is maintained in the system by using Berendsen thermostat \cite{berendsen84}. We then start cooling the system to $T=0$ for which a cooling rate of $10^{-3}/\tau_{LJ}$ time units is chosen for most of the results presented in the paper. To obtain better statistics, ensemble average is done over 200 independent initial configurations. However, to study the effect of cooling rate on NIO parameter, averaging is done over 50 ensembles.

\section{Results and Discussion}

\begin{figure}[!b]
	\centering
	\begin{tabular}{ccc}
		\includegraphics[width=0.3\linewidth]{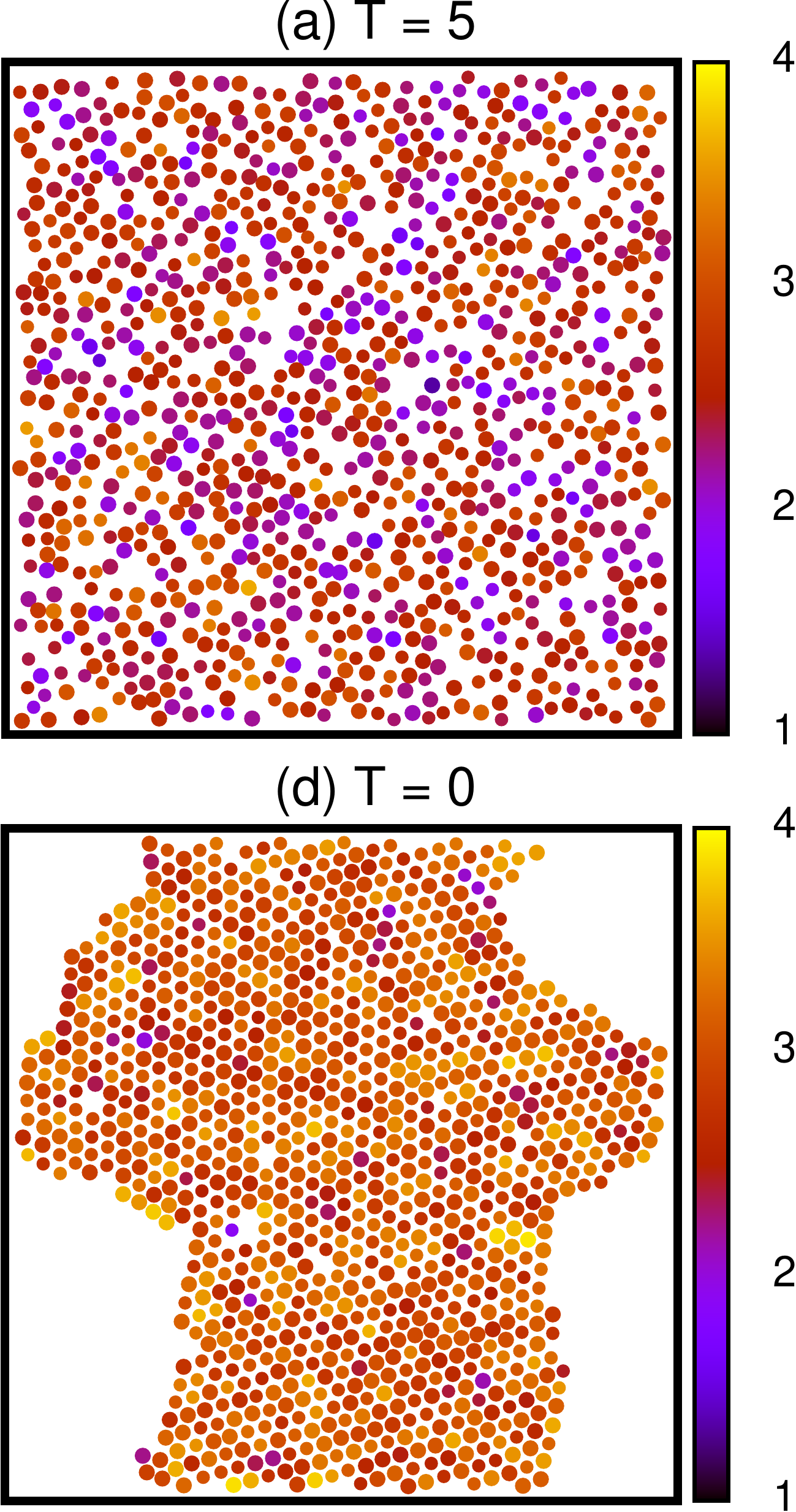} &
		\includegraphics[width=0.3\linewidth]{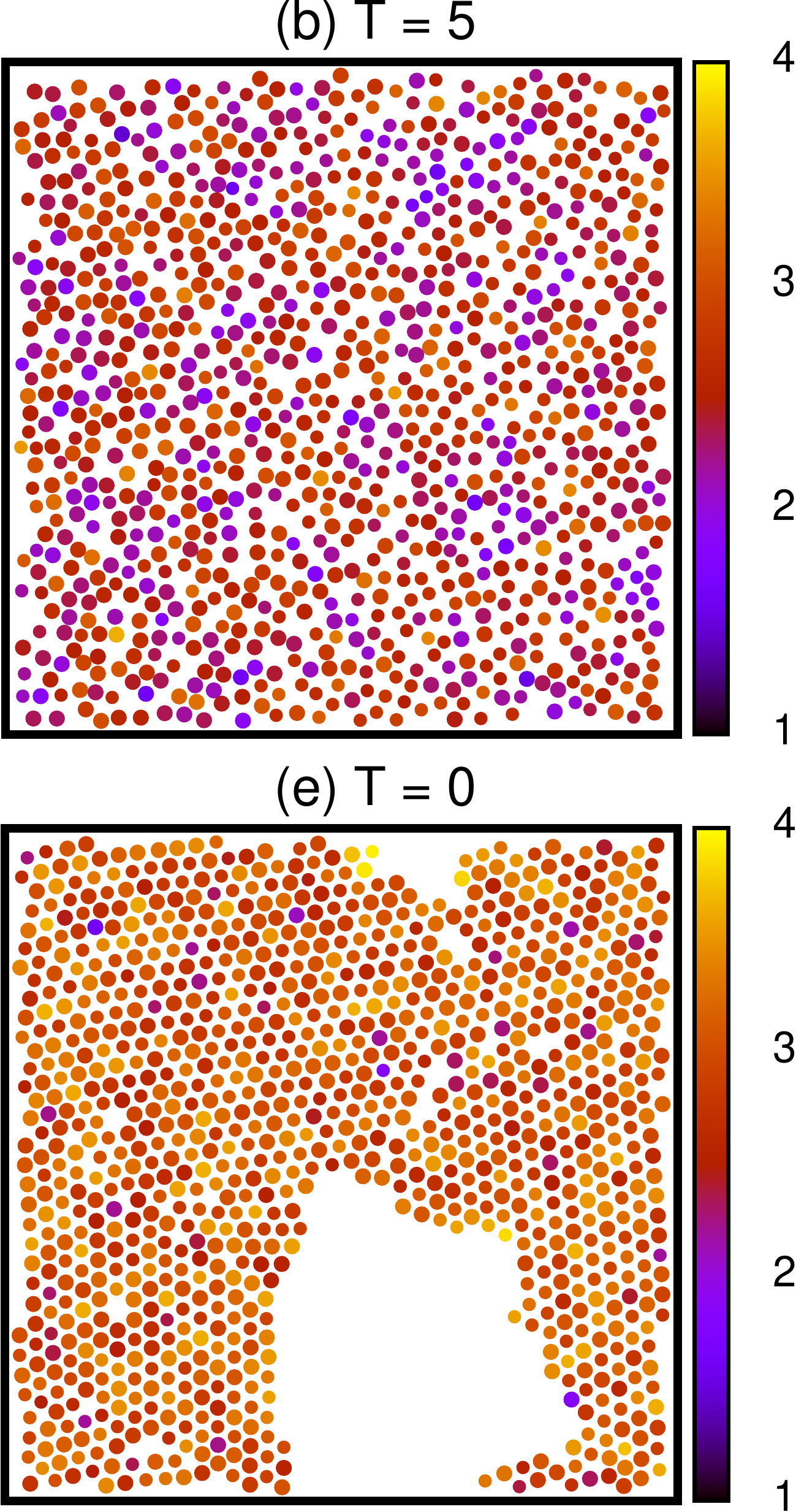} &
		\includegraphics[width=0.3\linewidth]{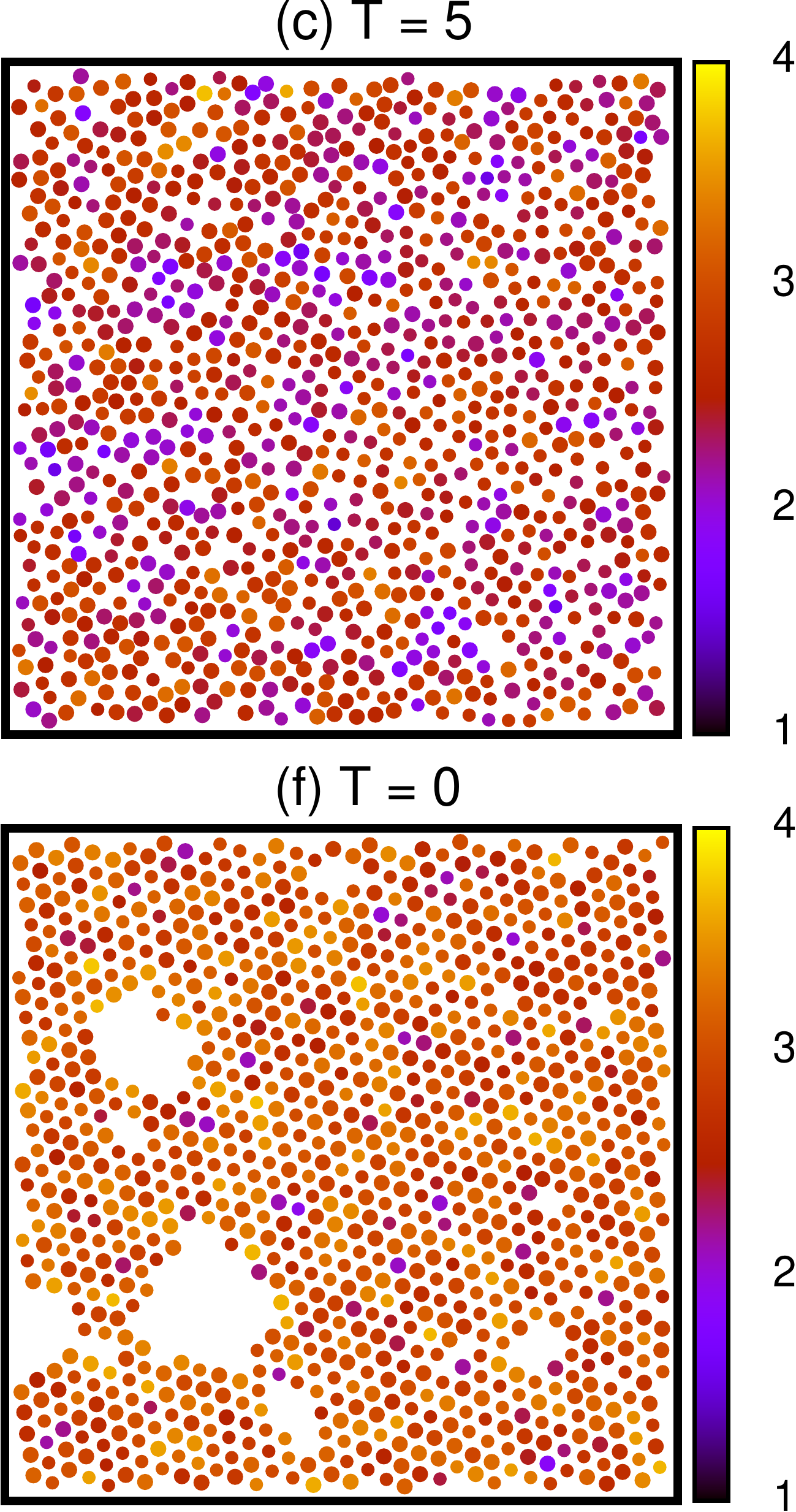}\\
	\end{tabular}
	\caption{Typical configuration of the bidisperse AID mixture is shown at temperatures $T=5$ (upper panel) and $T=0$ (lower panel) for concentration of A particles $x_A=0.5$ and size-ratio $\alpha=0.9$. The particles are colored according to their effective interaction parameter $\epsilon_{\rm eff}^{i}$ as displayed in the color bar. The parameters for which the configuration of particles is shown are (a) $\rho=0.7$, (b) $\rho=0.8$, (c) $\rho=0.9$, (d) $\rho=0.7$, (e) $\rho=0.8$, and (f) $\rho=0.9$.}
	\label{fig1}
\end{figure}

\begin{figure*}[!htbp]
	%\centering
	\begin{multicols}{1}
		\label{fig2a}
		\includegraphics[width=0.82\linewidth]{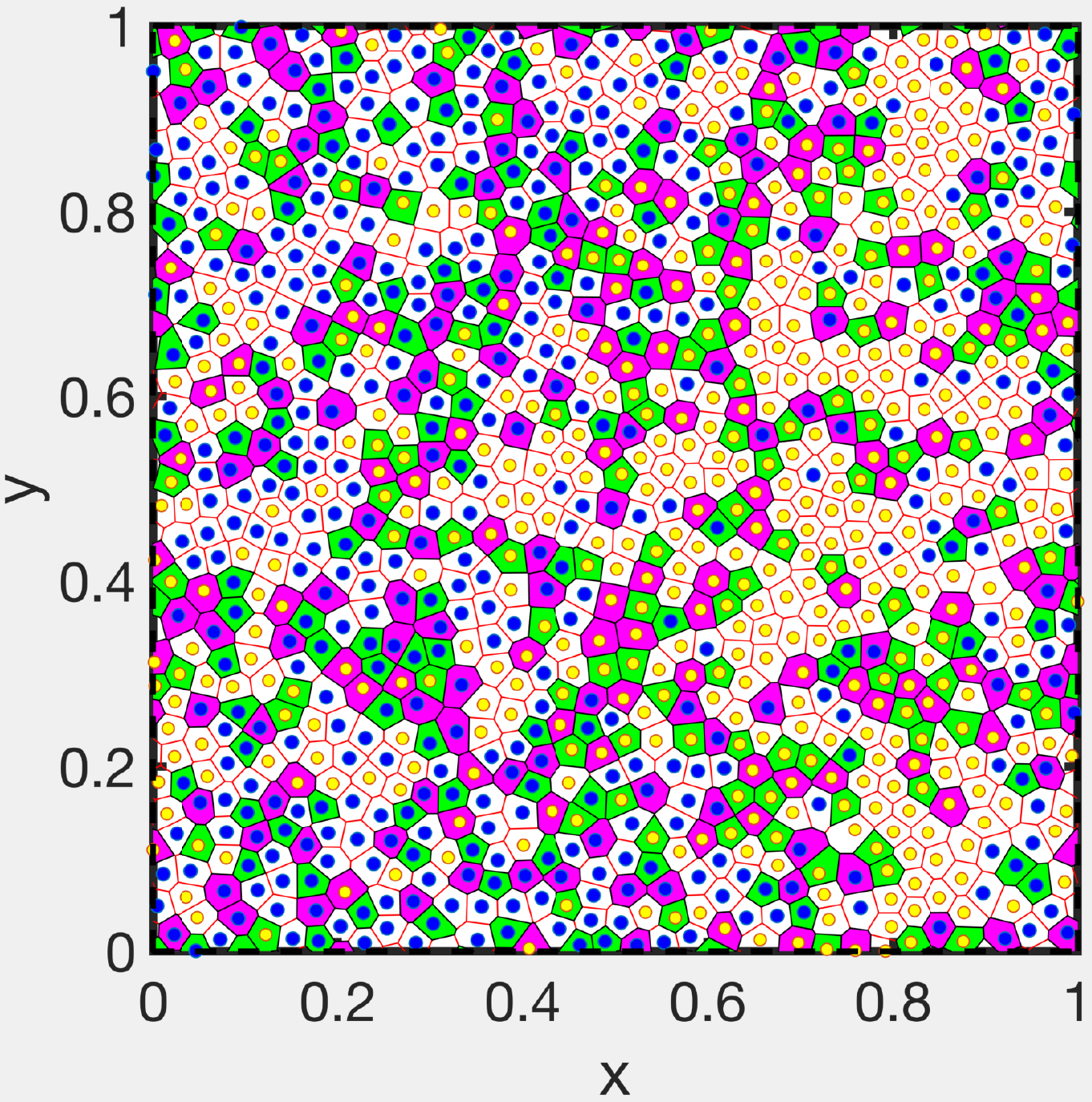}\par 
		\includegraphics[width=0.82\linewidth]{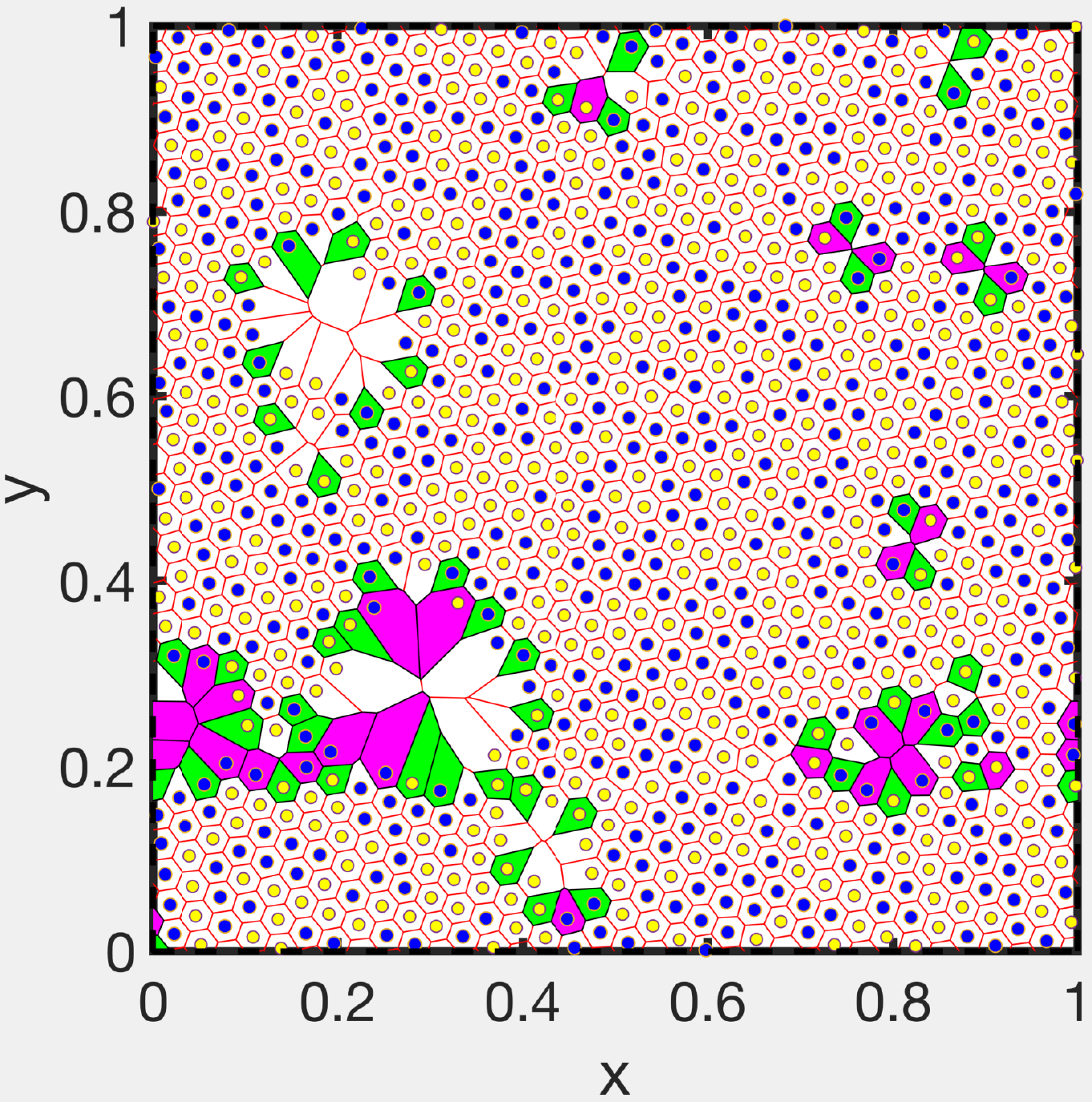}\par 
	\end{multicols}
	\caption{Voronoi tessellation of the snapshots for $\rho = 0.9$ at $T=5$ (left) and $T=0$ (right). Blue and yellow circles represent positions of A and B type particles, respectively.  Magenta cells enclose particles with 7 nearest neighbors and the green cells enclose particles with 5 nearest neighbors. White cells show mostly the crystalline patches of particles with 6 nearest neighbors.}
	\label{fig2}
\end{figure*}

As discussed in Sec. \ref{sec2} the bidisperse AID mixture is simulated at three densities $\rho = 0.7$, $0.8$ and $0.9$ at temperature $T=5$ and then quenched with cooling rate $10^{-3}/\tau_{LJ}$ to temperature $T=0$. Typical snapshots of the system in equilibrium at $T=5$ and $T=0$ are shown in Fig. \ref{fig1} for all three densities. The particles are colored according to their effective particle pair interaction parameter $\epsilon_{i}^{\rm eff}$ which is defined as,
\begin{equation}
\epsilon_{i}^{\rm eff} = \frac{1}{n_b}\sum_{j=1}^{n_{b}}\epsilon_{ij},
\end{equation}
where $n_b$ is number of neighbors of particles $i$ which are inside a cutoff radius of $1.7$. We observe that the system for which the particles were randomly distributed at $T=5$, as expected, has crystallized at $T=0$. The particles form domains of hexagonal lattice (also confirmed by Voronoi analysis in Fig.~\ref{fig2}), and voids which reduce as the density of system is increased. It can also be observed that the particles are organized according to their $\epsilon_i^{\rm eff}$ values. The particles with enhanced pair interaction parameter (PIP) $\epsilon_i^{\rm eff}$ have neighbors with enhanced PIP and the particles with smaller $\epsilon_i^{\rm eff}$ lie at the boundary. Such an order is called neighborhood identity ordering (NIO) and is in agreement with what has been reported previously~\cite{singh2019,shagolsem2015,shagolsem2016}. This NIO leads to increased cohesive forces between the particles.

Figure~\ref{fig2} displays the Voronoi analysis (one of the useful technique for studying the local atomic structure \cite{thsh18, dhj19}) for the snapshots corresponding to the density $\rho=0.9$ (shown in Figs.~\ref{fig1}(c) and \ref{fig1}(f)) for $T=5$ (left) and $T=0$ (right). The position of particles are represented by circles in blue (A-particles) and yellow (B-particles) colors. The colored boxes denote polygons containing particles with different number of neighbors. Green boxes represent pentagons enclosing particles with 5 neighbors. Magenta colored box encloses particles with 7 neighbors. Rest of the polygons are shown in white color. 
The configuration at $T=5$ has nearly $55.5\%$ hexagons, $22.8\%$ pentagons, and $18.3\%$ heptagons. Hence, the corresponding snapshot, of course,  looks more like a liquid. 
The quenched snapshot at $ T=0 $ has nearly $86.7\%$ hexagons, $8.6\%$ pentagons, and $2.5\%$ heptagons. This indicates that the quenched system crystallizes as it is mainly composed of hexagonal patches (in 2-dimension). This is due to Thue's theorem which states that the regular hexagonal packing is the densest of all possible circle packings in a 2-dimensional plane \cite{thue10}.

\begin{figure*}[!b]
	%\centering
	\begin{multicols}{1}
		\label{fig3a}
		\includegraphics[width=\textwidth]{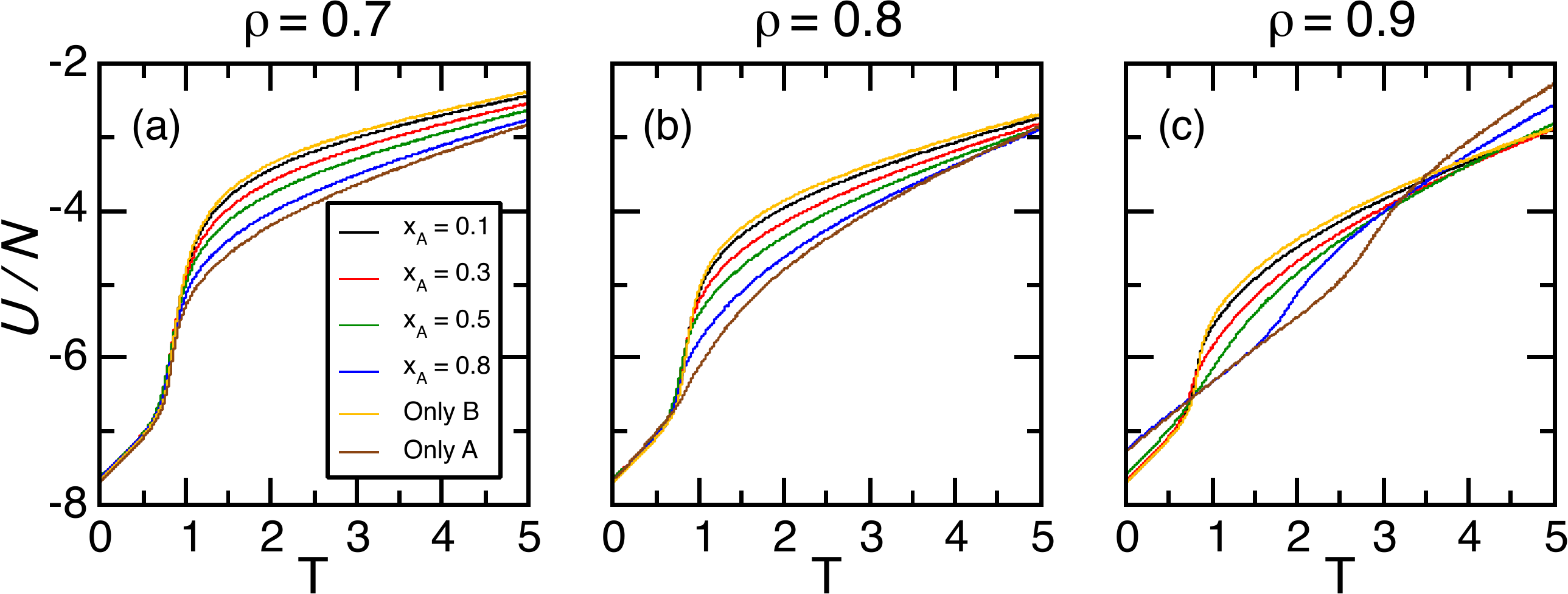}\par 
	\end{multicols}
	\begin{multicols}{1}
		\label{fig3b}
		\includegraphics[width=\textwidth]{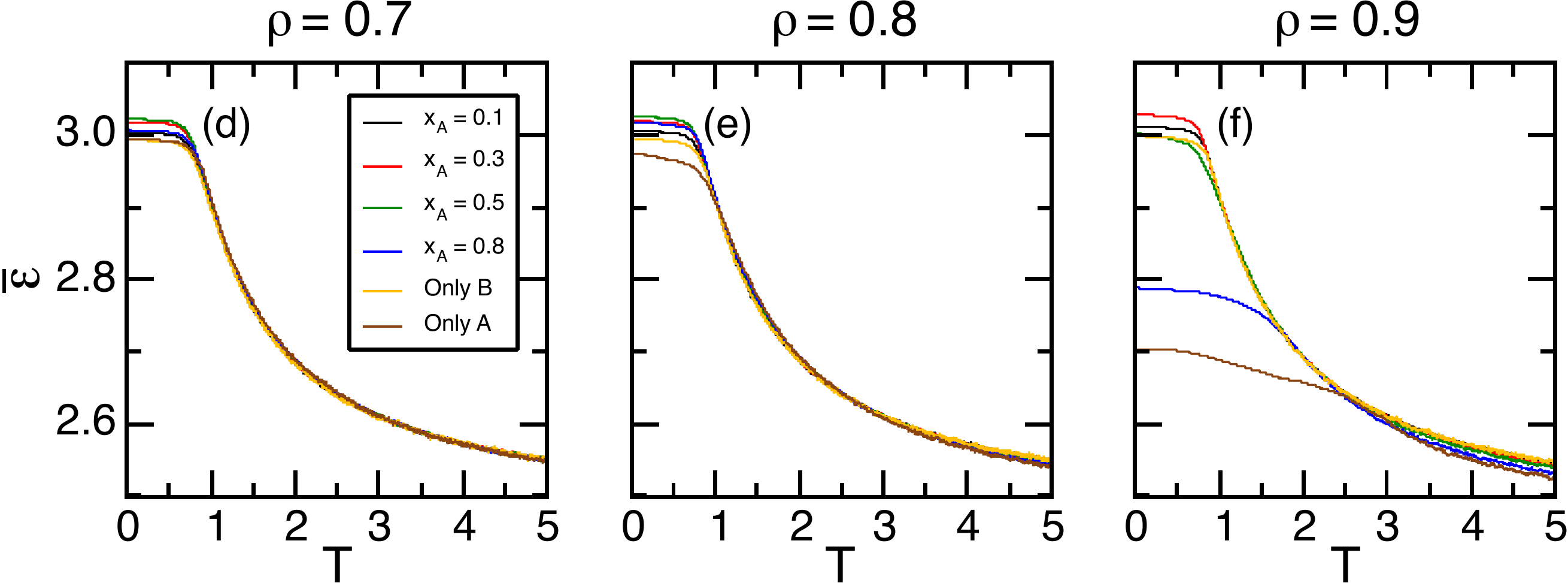}\par 
	\end{multicols}
	\caption{Upper panel: Dependence of potential energy per particle on temperature is shown for various concentration of the bigger particles at three different densities (a) $\rho=0.7$, (b) $\rho=0.8$ and (c) $\rho=0.9$. Lower panel: NIO parameter $\bar{\epsilon}$ is plotted against $T$ for same concentration of bigger particles as shown in the upper panel for same density values (d) $\rho=0.7$, (e) $\rho=0.8$ and (f) $ \rho=0.9$. For all these plots, size-ratio $\alpha$ is fixed at 0.9.}
	\label{fig3}
\end{figure*}

In the upper panel of Fig. \ref{fig3}, we plot the potential energy per particle $U/N$ with temperature as the system is quenched from $T=5$ to $T=0$. The plots are presented at densities $\rho = 0.7$, $0.8$, and $0.9$ for concentration of bigger particles $x_A = 0.1, 0.3, 0.5, 0.8$, $ 0.0\,(100\% B$-particles), and  $ 1.0\,(100\% A$-particles). It is observed that $U/N$ decreases with temperature as the system is cooled for all the densities. For $ \rho=0.7, 0.8 $, $U/N$ decreases with the increase in $ x_A $ for a fixed temperature (see Figs.~\ref{fig3}(a) and \ref{fig3}(b)). However, for $\rho=0.9$, this trend is reversed beyond the transition temperature at around $T=3$ as shown in Fig.~\ref{fig3}(c). This reverse trend cannot be observed for $ \rho=0.7$ and $0.8$ within the plotted temperature range (because we quench the system from $T=5$). The liquid to solid transition for $\rho=0.7$ in Fig.~\ref{fig3}(a) occurs at $T\approx 1.1$ for all the compositions. The transistion temperature for $\rho=0.6944$ for the same system at $x_A = 1$  is reported to be at $T\approx 1.1$ in~\cite{shagolsem2015}. The slope of the curve in fluid region increases as the value of $x_A$ is increased. This implies that the specific heat at constant volume is higher for higher $x_A$~\cite{singh2019}. As we increase the density, a gradual shift (increase) in the liquid to solid transition temperature with the increase in $x_A$ is observed as shown in Figs.~\ref{fig3}(b) and \ref{fig3}(c). For $\rho = 0.9$, the transition temperature increases from $T\approx 1.1$ for $x_A=0$ (only B) to $T\approx 3$ for $x_A=1$ (only A) in Fig.~\ref{fig3}(c). Note that A particles are bigger in size as compared to B particles.

It is already known that upon cooling, an AID system develops NIO ordering in the liquid state (close to crystallization)~\cite{singh2019,shagolsem2015}. The NIO in the system is characterized by the parameter $\bar{\epsilon} $  which is obtained by averaging over $\epsilon_i^{\rm eff}$ of all the particles. In the lower panel of Fig. \ref{fig3}, the NIO parameter $\bar{\epsilon}$ is plotted against the temperature. Note that the higher NIO parameter corresponds to lower potential energy per particle. Therefore, a decrease in $U/N$ while cooling the system implies the corresponding increase in $\bar{\epsilon}$ as can be confirmed by comparing the upper panel with the lower panel. The local organization of particles with higher $\epsilon_i^{\rm eff}$ around particles with higher $\epsilon_i^{\rm eff}$ leads to an increase in average PIP $\bar{\epsilon}$ of the system as the temperature is decreased. Below $T \sim 1.1$, $\bar{\epsilon}$ ceases to increase and saturates. The saturation value for different compositions varies as the density is increased. More specifically, this value is small for $x_A=1$ and as we start adding the B particles in the mixture, the saturation value increases and reaches up to a maximum value for $50:50$ mixture and then it again decreases (see Figs.~\ref{fig3}(d),~\ref{fig3}(e) and ~\ref{fig3}(f)). A similar phenomenon of increase in NIO parameter with different quenching protocols has been attributed to the mobility of defects below the liquid to solid transition temperature \cite{singh2019}. It is also observed that in the fluid region, $\bar{\epsilon}$ seems to be independent of density. However, at $T=0$, $\bar{\epsilon}$ varies with the densities. Furthermore, we observe that at low temperature, values of $U/N$ for different compositions are same. In contrast, a significant change is observed for $\bar{\epsilon}$ within the similar range of temperature. This finding suggests that $\bar{\epsilon}$ can be used as a good marker for differentiating systems with different values of $x_A$ at low temperature (below liquid to solid transition), which otherwise is not possible from the plots of $U/N$ for similar parameters. 
\begin{figure*}[!b]
	\begin{multicols}{1}
		\label{fig4a}
		\includegraphics[width=\textwidth]{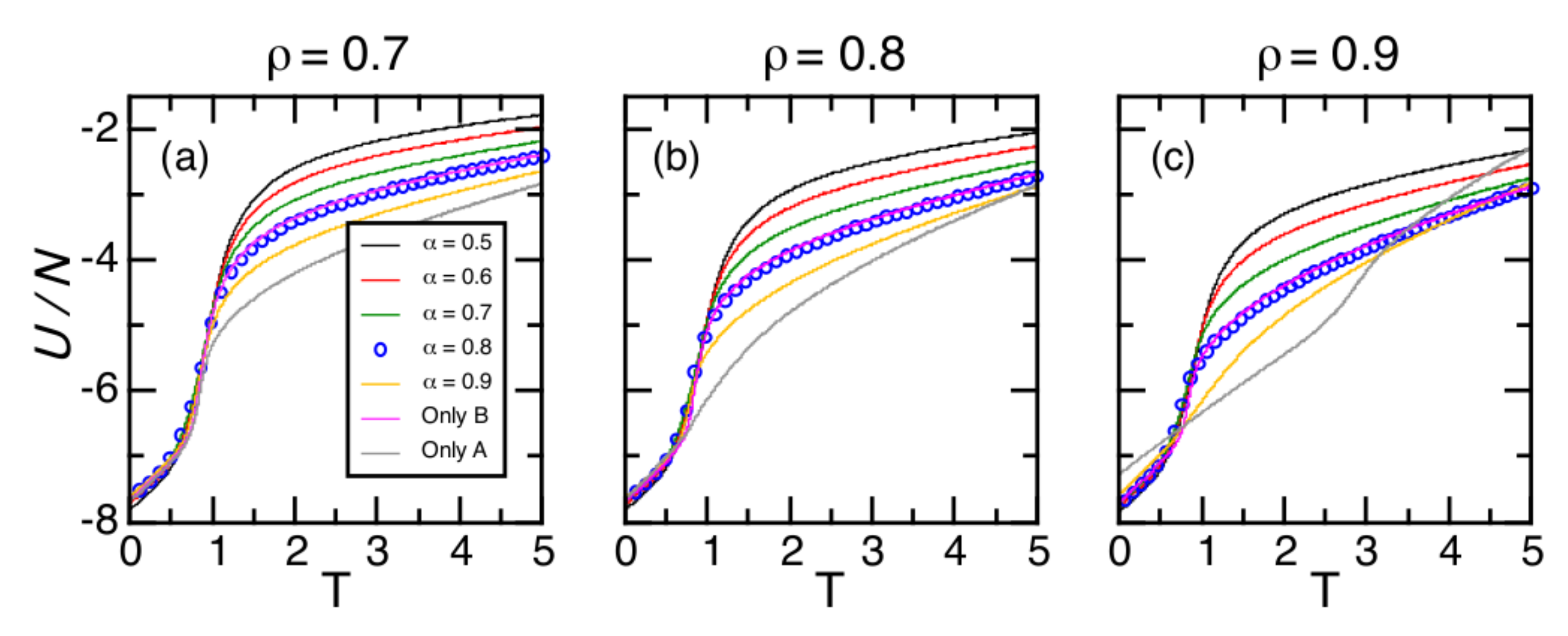}\par 
	\end{multicols}
	\vspace{-1cm}
	\begin{multicols}{1}
		\label{fig4b}
		\includegraphics[width=\textwidth]{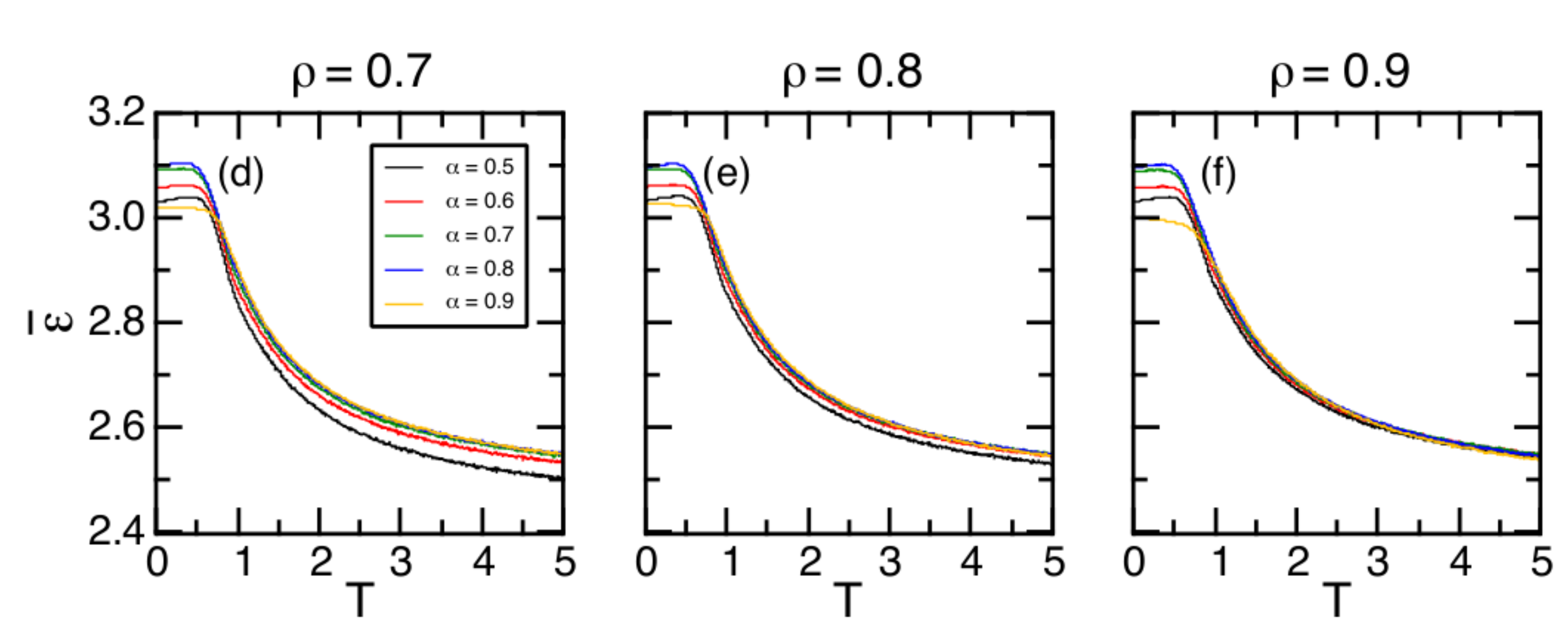}\par 
	\end{multicols}
	\caption{Upper panel: Temperature dependence of potential energy per particle is shown for size-ratios $\alpha = 0.5$, $0.6$, $0.7$, $0.8$, $0.9$ and for the cases where only particles of type A as well as type B are considered for densities (a) $\rho = 0.7$, (b) $\rho =0.8$, and (c) $\rho =0.9$. Lower panel: The change in NIO parameter $\bar{\epsilon}$ as the temperature is decreased with cooling rate $dT/dt = 10^{-3}/\tau_{LJ}$ is shown for same system parameters corresponding to Fig. \ref{fig4a} at densities  (a) $\rho = 0.7$, (b) $\rho =0.8$, and (c) $\rho =0.9$. For all these plots, concentration of A-type particles $x_A$ is fixed at 0.5.}
	\label{fig4}
\end{figure*}

Next, we study the potential energy per particle ($U/N$) and NIO parameter ($\bar{\epsilon}$) as a function of temperature for various values of size-ratio $\alpha$ ($=\sigma_B/\sigma_A$) and densities. The plots for $U/N$ vs. $T$ are shown in the upper panel of Fig.~\ref{fig4}, whereas in the lower panel of Fig.~\ref{fig4}, variation of $\bar{\epsilon}$ is plotted against $T$. We take $\alpha=0.5,0.6,0.7,0.8,0.9$ with $x_A=0.5$ ($50:50$ mixture) and one-component systems consisting of only A-type particles and only B-type particles as shown in the legend. When we increase $\alpha$, potential energy per particle is lowered in the fluid regime.  We also note that the curve for $\alpha=0.8$ (shown in blue circle) and for only B-type particles (shown in magenta line) are overlapping. This is because, for a system with  particles of only B-type ($\sigma=0.9$), the average size is 0.9 and for the case when $\alpha=0.9$, the average size is also 0.9. With the increase in density, the transition temperature slowly increases with $\alpha$, of course, the transition takes place over a broader range of temperature range as well. Similar to the case discussed in Fig.~\ref{fig3}, below the crystallization temperature, $U/N$ is not a good marker for comparing the systems with varying $\alpha$. However, $\bar{\epsilon}$ again turns out to be a good quantity to differentiate the curves with different values of $\alpha$ in the solid regime. This is clearly visible in the lower panel of Fig.~\ref{fig4}. The saturation value of $\bar{\epsilon}$  in the solid regime increases with $\alpha$ and reaches its maximum for $\alpha=0.8$. For $\alpha=0.9$ (yellow line), $\bar{\epsilon}$ is lower as compared to the case with $\alpha=0.5$. NIO parameter is minimum for the cases with only A-type as well as B-type particles (results not shown).

\begin{figure*}[!b]
	\begin{multicols}{1}
		\includegraphics[width=\textwidth]{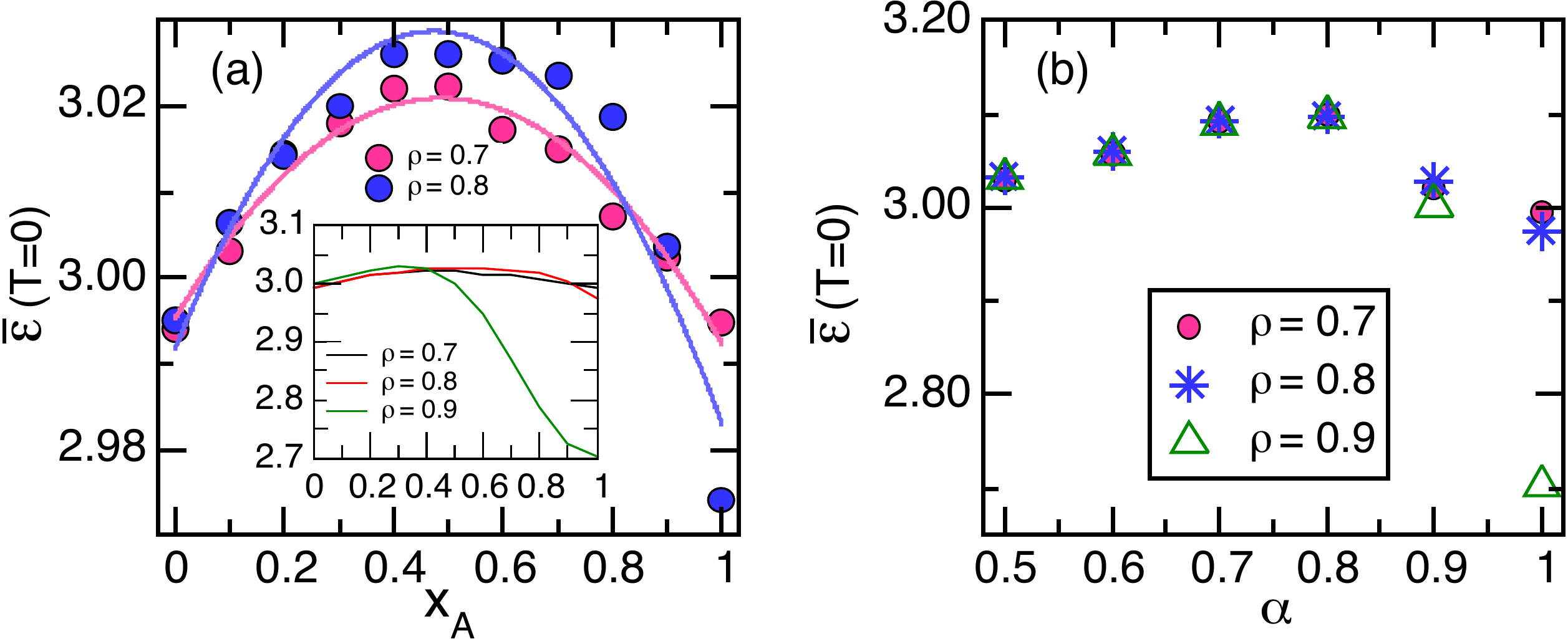}\par 
	\end{multicols}
	\caption{(a) Main: The variation of NIO parameter $\bar{\epsilon}$ with concentration of bigger particles $x_A$ is displayed at $T=0$  for $\rho =0.7$ and $0.8$. The solid lines are quadratic fit to the data and are symmetric about $x_A=0.5$. Inset: Plot of $\bar{\epsilon}(T=0)$ vs. $ x_A$ for $\rho=0.7,0.8,0.9$. A significant deviation is observed for $\rho=0.9$ as compared to quadratic and symmetric plots corresponding to $\rho=0.7,0.8$. The size-ratio is set at $\alpha=0.9$. (b) The dependence of $\bar{\epsilon}(T=0)$ on size-ratio $\alpha$ is shown at  densities $\rho =0.7$, $0.8$ and $0.9$. The value of $x_A$ is 0.5.}
	\label{fig5}
\end{figure*}
The dependence of NIO parameter $\bar{\epsilon}$ in the crystalline state, {\it i.e.}, at $T=0$ on concentration of bigger particles $x_A$ and size-ratio $\alpha$ is plotted in Fig.~\ref{fig5}. As $x_A$ is increased from $x_A = 0$ to $1$, it is observed that till $x_A = 0.5$, the value of $\bar{\epsilon}(T=0)$ increases monotonically and reaches a peak at $x_A = 0.5$. Upon increasing $x_A$ even further, $\bar{\epsilon}(T=0)$ starts decreasing. Therefore, NIO parameter can be increased by increasing the concentration of A particles until the $50:50$ ratio is achieved. For all the densities considered here, an increase of roughly $1.5\%$ is observed in the NIO parameter as compared to the systems with smaller particles only. If a comparison is made with a system of A particles only,  the increase in NIO parameter for $\rho=0.7$ and $0.8$ is again $~1.5\%$.  However, for $\rho = 0.9$, a significant increase of $~11\%$ is observed. A similar behavior of increase in $\bar{\epsilon}(T=0)$ was obtained by using the mechanism of defect mobility via invoking different quenching protocols \cite{singh2019}. Though, the plots of $\bar{\epsilon}(T=0)$ vs. $x_A$ for $\rho=0.7, 0.8$ exhibit a quadratic as well as symmetric behavior about $x_A=0.5$, the trend is absent for $\rho=0.9$ as shown in the inset of Fig.~\ref{fig5}(a). 
In Fig. \ref{fig5}(b), $\bar{\epsilon}(T=0)$ is plotted with size-ratio $\alpha$. It is observed that as the size-ratio is increased from $0.5$ to $0.8$, $\bar{\epsilon}(T=0)$ increases monotonically while for $\alpha \geq 0.9$, the NIO parameter  $\bar{\epsilon}(T=0)$ decreases and reaches a value lower than its value at other size-ratios. For densities $\rho =0.7$ and $0.8$, the NIO parameter increases upto $~3.5\%$ of its value for systems with A particles or B particles only. The density $\rho = 0.9$ displays the same increase as compared to a system of smaller particles only while an enhancement of $~15\%$ in the NIO parameter is observed on camparing it with a system with bigger particles only. 

\begin{figure*}[!h]
	\includegraphics[width=\textwidth]{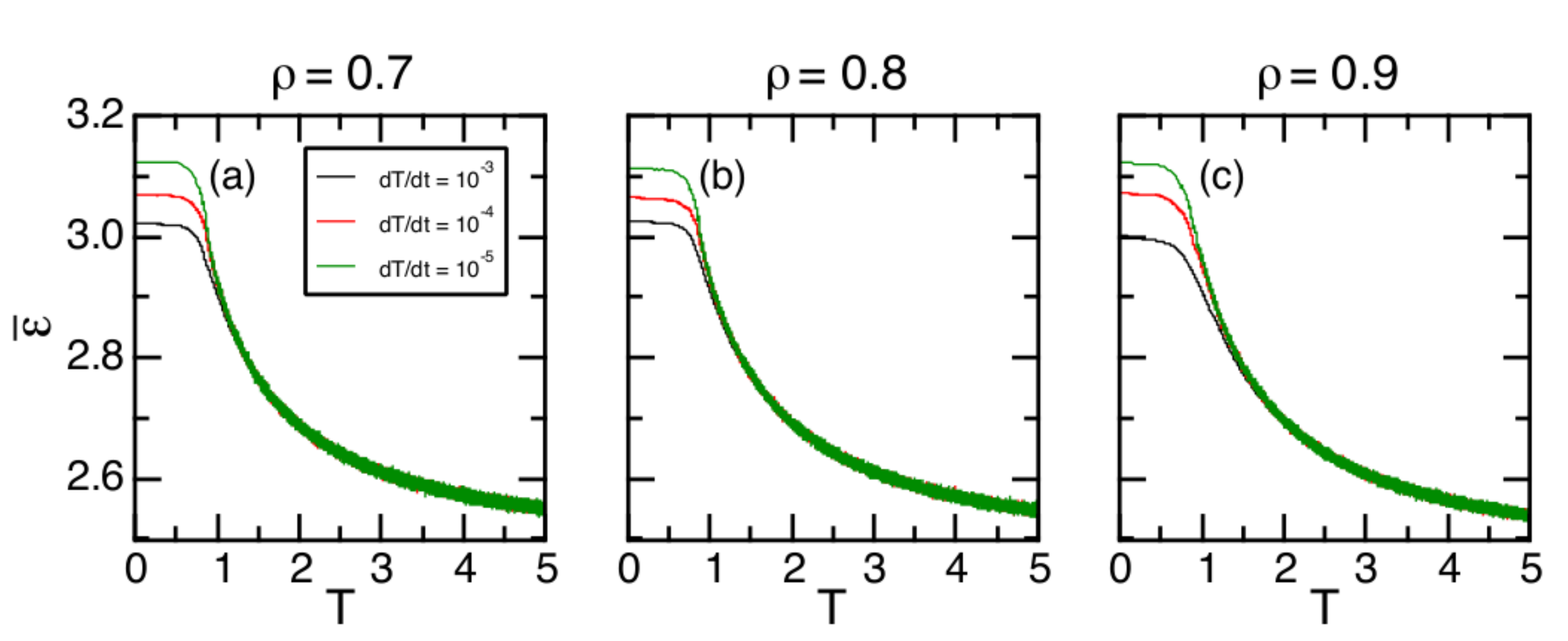}
	\caption{The neighborhood identity ordering parameter $\bar{\epsilon}$ is shown as a function of temperature for cooling rates $dT/dt = 10^{-3}/\tau_{LJ}$, $10^{-4}/\tau_{LJ}$ and $10^{-5}/\tau_{LJ}$ and densities (a) $\rho = 0.7$, (b) $\rho = 0.8$, and (c) $\rho = 0.9$. The other parameters are $x_A=0.5$ and $\alpha=0.9$.} 
	\label{fig6}
\end{figure*}

Finally, in Fig.~\ref{fig6}, we plot the dependence of NIO parameter $\bar{\epsilon}$ on temperature for $\rho =0.7,0.8,0.9$ and $x_A =0.5$ at lower cooling rates. It is observed that as the cooling rate is decreased,  $\bar{\epsilon}$ saturates to a higher value below the crystallization temperature. The slow quenching rate provides longer time for configurational sampling at each temperature and enhance the NIO parameter. This is an interesting finding and is analogous to a fluid being quenched (fast enough to avoid the first-order phase transition, i.e., crystallization transition) at different cooling rates and glasses with different mechanical properties are formed below glass transition temperature \cite{ediger2000}. In this context, a system with higher value of $\bar{\epsilon}$ would be analogous to a ductile glass (where system gets trapped in a deeper minima of the potential energy landscape \cite{ds01}), both being obtained at lower cooling rates. 

\section{Conclusion}
Designing materials with one or a few components poses a limitation in further improving the mechanical, optical, thermal, etc. properties. This restriction has led materials scientists as well as engineers to tailor polydisperse systems, e.g., multicomponent alloys and complex fluids, for improving these properties. We, therefore, perform atomistic simulations for a complex mixture which is polydisperse in energy and bidisperse in size. The polydispersity in energy is introduced by considering all pair interactions between the particles to be different. It is observed that mixing AID particles of two different sizes increase the attractive forces between them. The increased cohesive forces between the particles, in turn, enhances the mechanical strength of the system. The attractive forces between the particles are measured in terms of an NIO parameter, which shows a significant increase for specific compositions and size-ratios. An enhancement in the NIO parameter is also achieved by cooling the system to a crystalline state with reduced cooling rates. We report our findings for a wide range of parameters and thus, offer some guidelines to the materials scientists as to which parameter under certain conditions would be better suited to give the desired property. We trust that the present results will have further implications on the study of multicomponent metallic alloys and complex fluids.

\section*{Acknowledgments}
	Helpful discussion with Santosh Mogurampelly is gratefully acknowledged. M.P.'s research is supported by SERB-ECR grant No. ECR/2017/003091 from DST, India.

\section*{Conflict of interest}
The authors declare that they have no conflict of interest.

\end{document}